# How real is the random censorship model in medical studies?

Damjan Krstajic[1]

[1]Research Centre for Cheminformatics, Jasenova 7, 11030 Beograd, Serbia

Email address:  damjan.krstajic@rcc.org.rs

## Abstract

In survival analysis the random censorship model refers to censoring and survival times being independent of each other. It is one of the fundamental assumptions in the theory of survival analysis. We explain the reason for it being so ubiquitous, and we investigate its presence in medical studies. We differentiate two types of censoring in medical studies (dropout and administrative), and we explain their importance in examining the existence of the random censorship model. We show that in order to presume the random censorship model it is not enough to have a design study which conforms to it, but that one needs to provide evidence for its presence in the results. Blindly presuming the random censorship model might lead to the Kaplan-Meier estimator producing biased results, which might have serious consequences when estimating survival in medical studies.

## Introduction

The assumption that censoring times are independent of survival times is the basis for the majority of theory of survival analysis related to medicine. It is usually referred to as the random censorship model, or just random censoring. For example, Edward L. Kaplan and Paul Meier presume it in their paper [1]. Usually the assumption is briefly discussed at the beginning of textbooks. We can take the assumption as an axiom in the Aristotelian sense, i.e. as a statement worthy of acceptance and needed prior to learning anything, and hope that the large part of theory of survival analysis, which is based on it, is in accord with reality.

Here, however, we take a different approach. We first analyse the view of survival data as the competing risks of failure and censoring times, and then we explain its corollaries. Later we introduce two types of censoring, dropout and administrative. Furthermore, we use large sample simulation to show that the proportion of administrative censoring in our results depends on the length of the study, as well as on the length of the recruitment period.

Our main thesis is that it is not enough to have a design of the study which demonstrates the existence of the random censorship model, but that we also need to show that mechanism, upon whose existence we presumed the random censorship model, was in action.

## Methods

Our sample consists of subjects with failure times (uncensored) and subjects with observation times during which failure did not happen (censored). The i-th subject has either the failure time or the censoring time. We suppose that in the absence of censoring the i-th subject in the sample has failure time $T_i$, a random variable with survivor and density functions $F_i$ and $f_i$ respectively.

**Survival data as competing risks**

When introducing censoring one is usually presented with the view of survival data as the result of competing two risks: failure times (Ti) and censoring times (Ci).

$$T = \min(T_i, C_i)$$

$$\delta = I(T_i < C_i)$$

It is assumed that the censoring times Ci for the i-th subject is a random variable with survivor and density functions Gi and gi. In the *random censorship model* we assume that the censoring times Ci are stochastically independent of each other and of the independent failure times Ti [2]. It is worth pointing out that there isn't any statistical test for the independence between censoring and survival times nor could there be one [3]. *Non-informative censoring* is a special case of the random censorship model, where in addition to being independent, the distribution of survival times does not provide any information about the distribution of censoring times, and vice versa. Usually this is presented as F=F($\theta$) and G=G($\lambda$). *Informative censoring* is the case of random censorship model which is not non-informative.

There is a very good reason why the random censorship model has become so ubiquitous. It is called the *non-identifiability issue* [3][4][5]. It states that if we drop the assumption of independence it is impossible to consistently estimate the survivor function F. Fortunately, in medical studies we usually have random entry into the study, with fixed censoring date at the end of the study, which causes censoring to be independent of the failure times.

**Dropout and administrative censoring**

Unfortunately, not all censoring in medical studies are the result of subjects surviving the end of the study. There are cases where subjects drop out or they are withdrawn from the study. The cause of their censoring is different from the subjects who survived the end of the study. Therefore, there are two kinds of censored subjects in medical studies: dropout and administrative. In literature the dropout subjects are also referred to as "lost to follow-up".

If we accept that the random censorship model in medical studies exists due to random entry into the study with fixed censoring date at the end of the study, how can we presume that the dropout censoring times are independent of the failure times? In other words, if we cannot find an external mechanism which causes dropout censoring times to be independent of the failure times, how can we accept the random censorship model? It is again worth mentioning that we cannot have a statistical test for the independence between censoring and failure times to help us resolve this conundrum [3]. So we are left with two options. The first is to presume that dropout censoring times are independent of the failure times and blindly hope that our survival estimates are not biased. The second is to accept that dropout censoring time may not be independent of the failure times. In this paper we are interested in investigating the second option.

**Influence of competing risks in the random censorship model**

First we will show the dependency between the influences of competing risks in the random censorship model and the study length, as well as the relationship between the study length and the number of dropout censored subjects. We present two extreme cases. In both cases we conduct a study with random entry and fixed censoring date at the end of the study. The difference between the cases is the length of the study. In the first case we define a very long study time by the end of which all subjects would have either dropped out or had failures, i.e. without any administrative censored subjects. While in the second case we define a very short study time by the end of which we would end up with nearly all administrative censored subjects and hardly any failures and dropouts. In both cases there was obvious risk of censoring mechanism independent of the failure times, but in the first case it never happened in the study, while in the second it obviously did.

Just because we have random entry and fixed censoring does not mean that we can view the survival data as a competition of two independent risks, failure times and censoring times. We argue that we also need to check whether the results confirm that view. It is possible that one of the risks might never happen in reality, or its influence might be minute. The first case, with very long study length, is an example where mechanism, upon whose existence we presumed the random censorship model, was not in action at all. Therefore, in the first case, even though we had a random entry and fixed censoring data, we may not have the random censorship model.

## Simulations

In medical studies we don't know the distribution of the failure times, the distribution of administrative censoring times nor the distribution of dropout censoring times. However, we can calculate the proportion of them in our survival data. If we accept that the dropout censoring times may not be independent of failure times, then the KM estimator might be more biased as the proportion of dropout censored subjects grows. We will use large sample simulations to demonstrate this.

Here we will describe several scenarios which in our opinion mimic real-life situations in medicine. We first generated failure times Ti (i=1..N) for N subjects as random samples from a survival distribution F and then we executed censoring according to the scenario. In scenarios where it was possible, we were interested to see what happens when various proportions P were being censored.

We used the KM estimator to generate survival curves for each simulation. We calculated the estimated median survival time for each simulated survival data. In addition, we used the actual median survival time, i.e. the median survival time of the true distribution, as a fixed time point. The probability of survival at the actual median survival time is 0.5 for the dataset without censored samples. We were interested to see how the KM estimator would calculate the probability of survival in each simulation at the actual median survival time and to see how much it would differ from 0.5.

We simulated the situation when a dropout event is dependent of the failure times. We presumed that the distribution of the dropout times D is the compound distribution of a distribution X and the distribution of failure times F, i.e. D=X o F . We were interested to see what happens when the survival data consists of different proportions of dropout subjects.

We presumed that administrative censored subjects have censoring times with the distribution of study observation times A, which is non-informative and independent of F.

**Scenario 1 – by the time study ended all subjects have either dropped out or had failures**

This is a scenario where we were in a position to observe all possible failure times. This means that either there isn't any end study defined, or by the time the study ended all subjects have either dropped out or had failures. There is a proportion P of subjects which have dropped out of the study. Here is the algorithm for the scenario 1.

1. Input parameters are (N, P, X).
2. Randomly select P proportion from N subjects to be dropout censored subjects.
3. For each selected censored subject assign its last known survival time to be a number equal to a random probability of distribution X multiplied by its actual failure time.

**Scenario 2 – without dropout and random entry to study**

This is a scenario where we have beginning and the end of a study and the recruitment is performed during the whole time of the study (Tstudy). This means that random entry to study may happen during whole time of the study. Subjects are not allowed to leave the study and, therefore, there aren't any dropout censoring subjects. Here is the algorithm for the scenario 2.

1. Input parameters are (N, Tstudy).

2. For each subject assign its study observation time Ei as a random number between 0 and Tstudy. The study observation time for each subject is the time between its entry to study and the end of the study.

3. For each subject calculate (Ti – Ei).

    a) If (Ti – Ei) > 0 then the i-th subject survived the end of study. In that case, the i-th subject is then censored with its last known survival time equal to Ei, i.e. Ci=Ei

    b) If (Ti – Ei) <= 0 then the i-th subject had a failure during the study. Therefore, its failure time is then unchanged.

**Scenario 3 – without dropout and random entry to study during recruitment time**

This is a scenario where we have beginning and the end of a study and the recruitment time (Trecruitment) is shorter than the study time (Tstudy). This means that random entry to study may happen only during the recruitment time. The algorithm for the scenario 3 differs from the scenario 2 only in the first two steps.

1. Input parameters are (N, Tstudy, Trecruitment).
2. For each subject assign its study observation time Ei as a random number between (Tstudy-Trecruitment) and Tstudy.

**Scenario 4 – with dropout and and random entry to study during recruitment time**

The same as is in scenario 3, here we have beginning and the end of a study and random entry to study may happen only during the recruitment time. However, there will be a proportion P of subjects which will drop out if the study did not have an end date. Again we presume that the dropout times have compound distribution of X and F, i.e. D = X o F. Here is the algorithm for the scenario 4.

1. Input parameters are (N, P, X, Tstudy, Trecruitment).
2. Randomly select P proportion from N subjects to be dropout censored subjects.
3. For each selected censored subject assign its last known survival time to be a number equal to a random probability of distribution X multiplied by its actual failure time.
4. For each subject assign its study observation time Ei as a random number between (Tstudy-Trecruitment) and Tstudy.
5. For each subject calculate ($T_i - E_i$), where Ti is either the failure time or its last know survival time.
   a) If ($T_i - E_i$) > 0 then the i-th subject survived the end of study. In that case, the i-th subject is then censored with its last known survival time equal to Ei, i.e. Ci=Ei
   b) If ($T_i - E_i$) <= 0 then the i-th subject had a failure or was dropped out during the study. Therefore, its failure time or last known survival time is then unchanged.

## Results

In our simulations we used N = $10^7$ as a relatively large number for sample size. We executed simulations with larger numbers like $10^8$ and $10^9$, and we got almost identical results (not published). We simulated the failure times with an exponential distribution with an input parameter λ. The median survival time is $\ln(2)/\lambda$, and we refer it to as the actual median survival time.

As regards the dropout events we were interested to examine the following three options:

a) dropout happens more likely after the entry to study than closer to the failure time
b) dropout happens with equal chances at any time between 0 and the failure time
c) dropout happens more likely closer to the failure time than after the entry to study

Therefore, we chose the following three distributions for dropout events:

a) compound distribution of the beta distribution (alpha=2. beta=5) and the distribution of failure times F
b) compound distribution of the uniform and the distribution of failure times F
c) compound distribution of the beta distribution (alpha=2. beta=5) and the distribution of failure times F

When simulating administrative censoring we were interested in two situations:

a) random entry during the whole time of the study
b) random entry during the recruitment period

In both cases we presumed administrative censoring times to have the uniform distribution, but with different lengths. In the first case the administrative censoring time may be any time between 0 and Tstudy, while in the second case it may be any time between (Tstudy-Trecruitment) and Tstudy. When simulating different study lengths and recruitment periods we used the actual median survival time as a unit of time.

**Scenario 1 simulation results**

Figures Fig1a, Fig1b and Fig1c show bias generated in scenario 1 with three different dependent censoring times when 20%, 40% and 60% of data are censored. In Table 1 we present for each simulation the estimated survival at the actual median survival time and the ratio between the estimated survival time and the actual median survival time. In Figure 2 we show a KM graph for a simulation with 40% data censored with three different dependent censoring times.

**Scenario 2 simulation results**

In Table 2 we show percentage of administrative censoring in scenario 2 for various study lengths, as well as the estimated survival at the actual median survival time. The scenario 2 is an example of the random censoring model, and as expected, the estimated survivals at the actual median survival time are 0.5. In Figure 3 we show a KM graph for a simulation where study length is 3 ams (actual median survival), which as the result caused 42.15% of data to be censored.

**Scenario 3 simulation results**

In Table 3 we show percentage of administrative censoring in scenario 3 for various combinations of study lengths and recruitment periods. The scenario 3 is again an example of the random censoring model, and as expected, the estimated survivals at the actual median survival time are 0.5.

**Scenario 4 simulation results**

Scenario 4 is the most complex of all scenarios as it contains dropout and administrative censoring subjects. It is a combination of scenarios 1 and 3. We fixed recruitment period and study length to be one half and three actual median survival times respectively. The same as in scenario 1 we simulated three dependent censoring times with P taking values 10%, 20% and 30%. In Table 4 we present for each simulation the estimated survival at the actual median survival time and the ratio between the estimated survival time and the actual median survival time, as well as the percentage of dropout and administrative censored subjects in each simulation.

## Discussion

Our simulations for scenarios 1 and 4 confirm the fact that the more censored subject we have whose censoring times are dependent of the failure times, the more biased our KM estimator would be. Furthermore, the extent of bias depends on the type of dependency between dropout and failure times. It is not the same if dropouts tend to happen earlier rather than later in the study.

As expected, our simulations confirm that the longer the study the fewer administrative censored subjects will be in the survival data. Equally we can say that the longer the recruitment period the more administrative subjects there will be in our sample. Therefore, as regards medical studies, where we have random entry and fixed censoring date at the end of the study, we may conclude that the length of a study and length of the recruitment period affect the influence of risks in the random censorship model.

Without any disregard to numerous authors who have significantly contributed to this field of research, we think that two major references mentioned so far deserve special attention and additional discussion.

- In 1974 Anastasios Tsiatis published a short and seminal paper on the non-identifiability issue [4]. He has shown that in the case of multiple competing risks, if they are not mutually independent, the model of potential survival times is unidentifiable. This means that as we view the survival data as the competition of two risks (failure times and censoring times) then if the censoring times are dependent of the failure times then we wouldn't be able to estimate survival. Figures 2 and 3 show what this means in practice. In Figure 3 we show a KM survival curve with 42.15% censored data where censoring times are independent of the failure times. Obviously in this case the KM estimator is unbiased, and our large sample simulation confirms the asymptotic of the KM estimator in the random censorship model [6]. However, in Figure 2 we show KM survival curves of the same sample with 40% censored data but with censoring times dependent of the failure times. Obviously with dependent censoring we cannot use the KM estimator. Figures 2 and 3 are our examples of the non-identifiability issue.
- In 1975 Arthur V Peterson published a technical report on the subject of non-parametric estimation in the competing risks problem [3]. His paper was primarily concerned with properties of the KM estimator in the competing risks problem. In the last section of his paper he considers the competing risks problem when the condition of independence of the risks is dropped, and he has proved the theorem (8.4.2) which states: *The observations in the competing risks problem, and hence any statistics based on them, contain no* *information on whether or not the risks are independent, and contain no information on whether the necessary and sufficient conditions for consistency of the Kaplan-Meier estimator is met*. The corollary is not only that estimating survival with a KM estimator when the risks are dependent might be biased, but also that we cannot create a statistical test for the independence between the competing risks. Suppose that such statistical test exists then we would be able to compare a survival data which has dependent censoring with the one which has independent censoring. This would mean that we would be able to identify a survival data with the independent censoring which is in contradiction with the non-identifiability of the independent censoring. Therefore, we cannot have the statistical test for the independence between the competing risks.

It is worth remembering that with a KM estimator we are estimating the survival curve. Therefore, the longer the study the more failures we will be able to observe, which as the consequence would mean that the KM estimation would have less variance. However, the longer the study the more dropouts may be in our sample, which again would lead to the KM estimation being potentially more biased. As we have shown, we cannot know the extent of the bias as we don't whether and how the dropout times are dependent of the failure times. So how should we proceed if there are dropouts in our sample? It is difficult to answer this question because the vast majority of published survival data do not differentiate between dropout and administrative censoring subjects. Therefore, we don't have enough past examples nor previous research upon which we could base our advice. Nevertheless, to start with we would suggest comparing the KM estimate based on all subjects (less variance and potentially biased) and the KM estimate using only failures and administrative censored subjects, i.e. without dropouts (more variance and no bias).

We argue that prior to accepting the random censorship model, not only that we need to specify which factors exist in reality which cause the censoring times to be independent of the failure times, we also need to show that they did affect our results. A good example would be scenario 3. The mechanism which creates independent censoring times definitely exists, but its influence on our results diminishes as we extend the study length and shorten the recruitment period, Therefore, in addition to acknowledging the existence of such mechanism in the design of the study, we also suggest to assess the proportion of administrative censored subjects in the survival data prior to accepting the random censorship model.

If we accept the view of the survival data in medical studies as consisting of failure times ($T_i$), dropout censoring times ($D_i$) and administrative censoring times ($A_i$), where the administrative censoring times are independent and non-informative of the failure times, while the dropout censoring times may not be independent of the failure times, we may view the survival data as the competition of three risks, i.e.

$$T = \min(T_i, D_i, A_i)$$

We think that survival data in medical studies ought to be viewed as the competition of three risks instead of two. Dropouts are not inevitable, but they should be expected. One of the points in the Nuremberg Code [7] states that the human subjects must be free to immediately quit the experiment at any point when they feel physically or mentally unable to go on.

The issue of potential bias, caused by existing dropouts, may become serious when comparing two survival curves in a randomised clinical trial. If, for example, in one treatment arm we have dropouts while in another none, how should we then compare them? We are not aware of any published solution. The best advice we were able to find is not to come to that situation, or to avoid it as much as possible. Double-sampling [8][9][10][11][12] is a design that accepts that dropouts are inevitable and selects a subset of them and devotes enough resources to find missing information about their survival.

We are not the first to introduce dropout and administrative censored subjects [13]. Nor the fact that dependent censoring may produce biased KM estimates is anything new [14]. There were also others who performed simulations to question how much dropouts is too much [15]. However, as far as we are aware, nobody yet has suggested to view the survival data as the competition of three risks. By acknowledging the existence of the dropout as a separate risk, we can work on reducing or removing bias which they may generate.

Scenarios 1, 2 and 3 are examples of extreme cases where censoring times are either all dependent or all independent of the failure times. In our opinion, the reality is somewhere in between, like scenario 4, and the problem is that the majority of theory of survival analysis related to medicine is only concerned with one extreme case.

## Conclusion

We think that the random censorship model is presumed too often without providing case for its existence in reality. We have shown that prior to accepting the random censorship model we need to produce evidence for its existence in the design of the study as well as in the results. The fact that the non-identifiability issue looms over the competing risk presentation of the survival data should not be the reason for blindly sticking to the random censorship model. On the contrary, we think that it should be an impetus to create new ways of viewing and managing survival data.

## Acknowledgments

The author would like to thank dr Ljubomir Buturovic, for his comments and suggestions on an earlier draft of the paper.

## References

1. Kaplan EL, Meier P. Nonparametric estimation from incomplete observations. Journal of the American statistical association. 1958 Jun 1;53(282):457-81.
2. Kalbfleisch JD, Prentice RL. The statistical analysis of failure time data. John Wiley & Sons; 2011 Jan 25.
3. Peterson AV. *Nonparametric estimation in the competing risks problem* (Doctoral dissertation, Dept. of Statistics, Stanford University).
4. Tsiatis A. A nonidentifiability aspect of the problem of competing risks. Proceedings of the National Academy of Sciences. 1975 Jan 1;72(1):20-2.
5. Tsiatis AA. An example of nonidentifiability in competing risks. Scandinavian Actuarial Journal. 1978 Oct 1;1978(4):235-9.

6. Breslow N, Crowley J. A large sample study of the life table and product limit estimates under random censorship. The Annals of Statistics. 1974;2(3):437-53.

7. Code N. The Nuremberg Code. Trials of war criminals before the Nuremberg military tribunals under control council law. 1949(10):181-2.

8. Frangakis CE, Rubin DB. Addressing an Idiosyncrasy in Estimating Survival Curves Using Double Sampling in the Presence of Self-Selected Right Censoring. Biometrics. 2001 Jun 1;57(2):333-42.

9. Baker SG, Wax Y, Patterson BH. Regression analysis of grouped survival data: informative censoring and double sampling. Biometrics. 1993 Jun 1:379-89.

10. Baker SG. Discussion of double sampling for survival analysis. Biometrics. 2001 Jun 1;57(2):348-50.

11. An MW, Frangakis CE, Musick BS, Yiannoutsos CT. The Need for Double-Sampling Designs in Survival Studies: An Application to Monitor PEPFAR. Biometrics. 2009 Mar 1;65(1):301-6.

12. An MW, Frangakis CE, Yiannoutsos CT. Choosing profile double-sampling designs for survival estimation with application to President's Emergency Plan for AIDS Relief evaluation. Statistics in medicine. 2014 May 30;33(12):2017-29.

13. Fisher LL, Kanarek PA. Presenting censored survival data when censoring and survival times may not be independent. Reliability and Biometry. 1974:303-26.

14. Klein JP, Moeschberger ML. Asymptotic bias of the product limit estimator under dependent competing risks. Indian Journal of Productivity, Reliability and Quality Control. 1984;9:1-7.

15. Kristman V, Manno M, Côté P. Loss to follow-up in cohort studies: how much is too much?. European journal of epidemiology. 2004 Aug 1;19(8):751-60.

# Figures

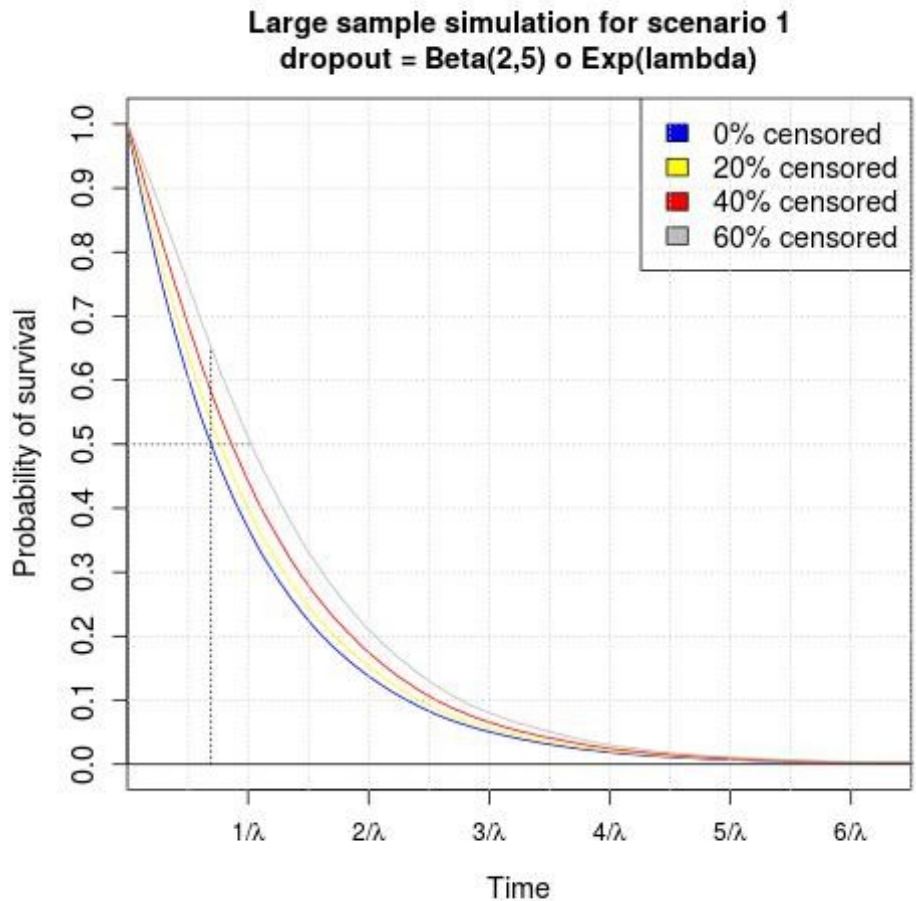

**Figure 1a - Large sample simulation for scenario 1 a**
Large sample simulation for scenario 1 where dropout distribution is compound distribution of the beta distribution (alpha=2. beta=5) and the distribution of failure times (exponential distribution)

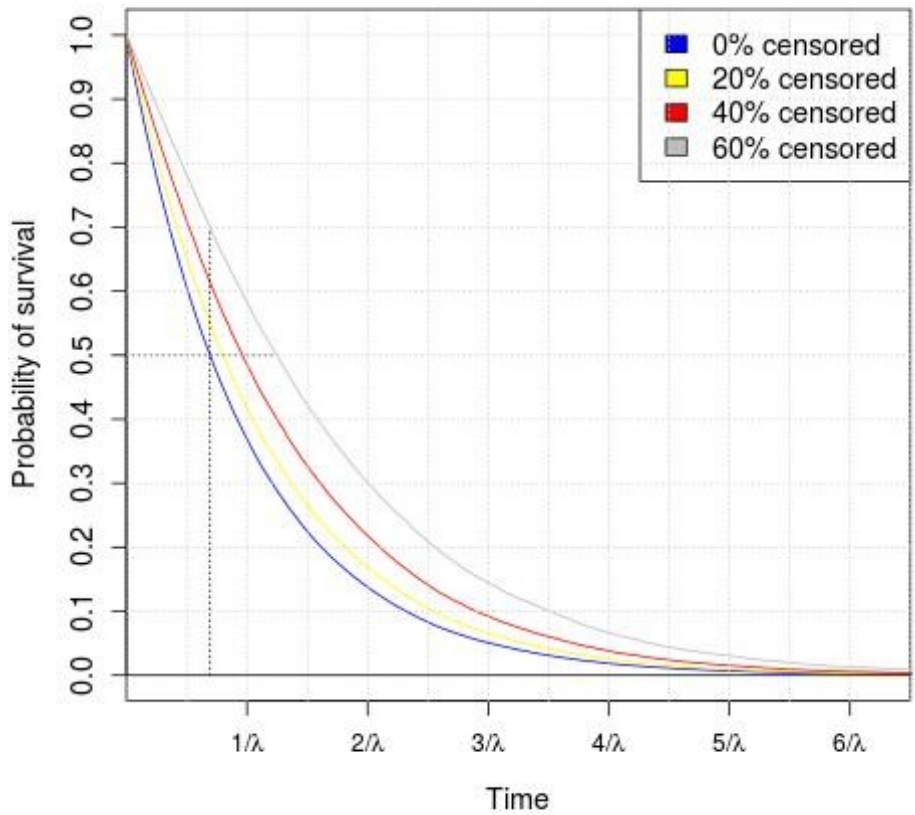

**Figure 1b - Large sample simulation for scenario 1 b**

Large sample simulation for scenario 1 where dropout distribution is compound distribution of the uniform distribution and the distribution of failure times (exponential distribution)

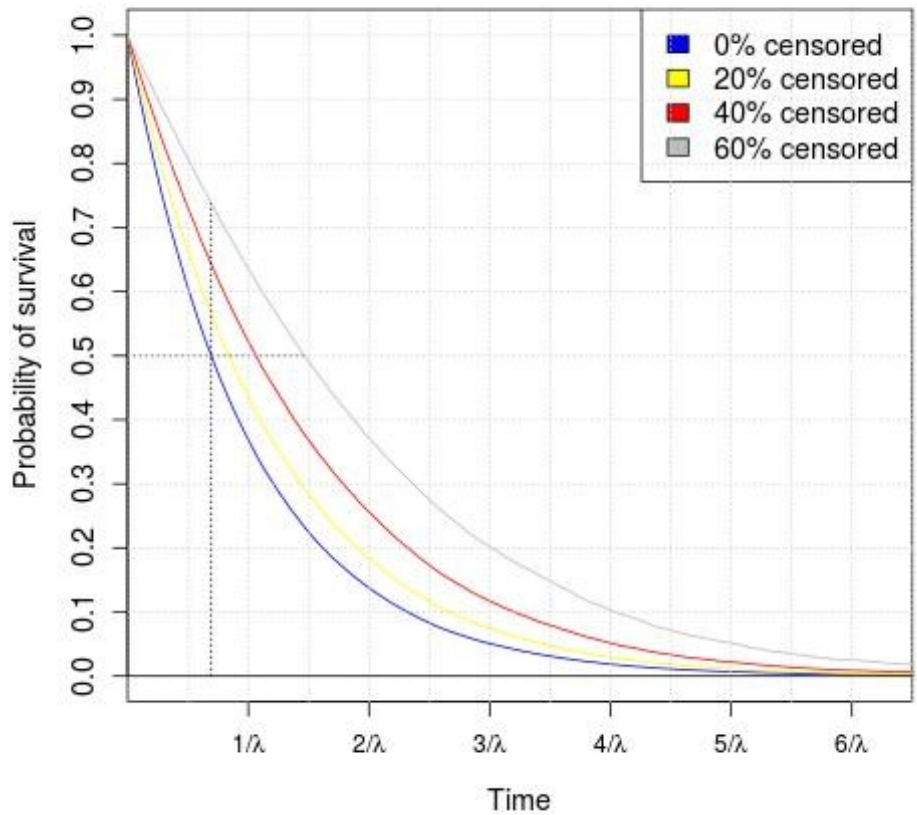

**Figure 1c - Large sample simulation for scenario 1 c**

Large sample simulation for scenario 1 where dropout distribution is compound distribution of the beta distribution (alpha=5. beta=2) and the distribution of failure times (exponential distribution)

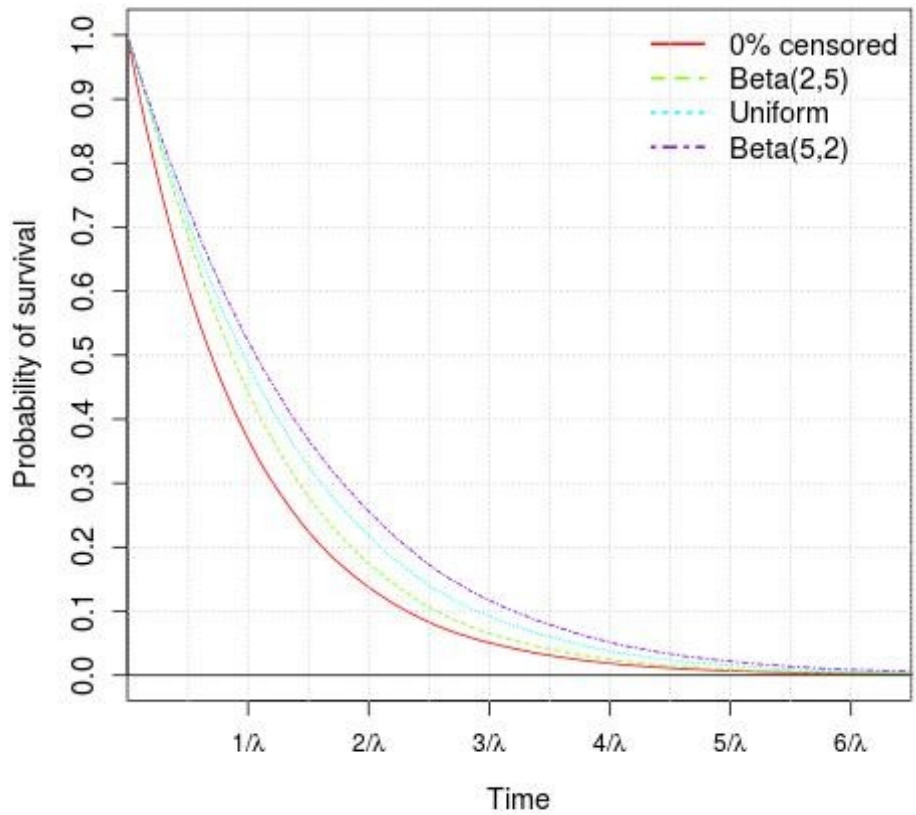

**Figure 2 - Large sample simulation for scenario 1 with 40% censored**

Large sample simulation for scenario 1 with 40% censored and three different dropout distributions, all dependent of the distribution of failure times (exponential distribution)

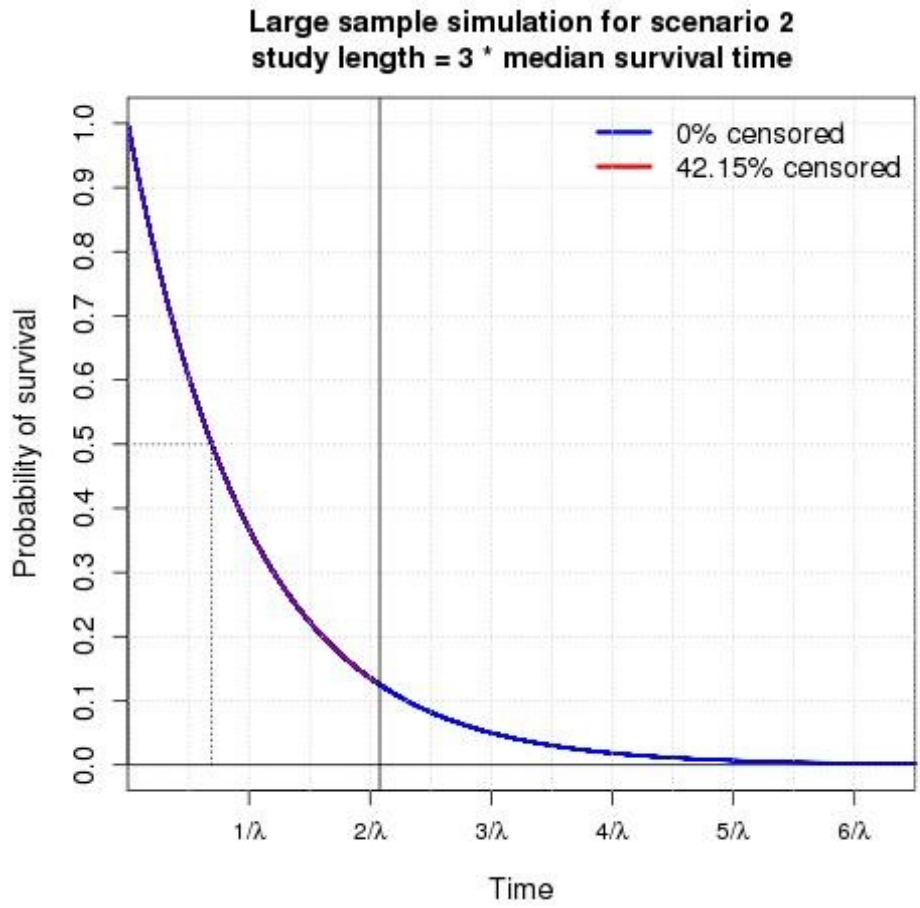

**Figure 3 - Large sample simulation for scenario 2 with 42.15% censored**
Large sample simulation for scenario 2 where study length is equal to three median survival times. The recruitment, i.e. entry to study, is performed during the whole time of the study.

# Tables

**Table 1 - Scenario 1**

The estimated survival at the actual median survival time (S(ams)), and the ratio between the estimated median survival time and the actual median survival time (ms / ams) are shown for each simulation.

|  | Dropout distribution | % censored | S(ams) | ms / ams |
|---|---|---|---|---|
| Scenario 1a | Beta(5,2) o F | 20 | 0.535 | 1.103 |
| Scenario 1a | Beta(5,2) o F | 40 | 0.582 | 1.246 |
| Scenario 1a | Beta(5,2) o F | 60 | 0.651 | 1.473 |
| Scenario 1b | Uniform o F | 20 | 0.55 | 1.155 |
| Scenario 1b | Uniform o F | 40 | 0.614 | 1.384 |
| Scenario 1b | Uniform o F | 60 | 0.7 | 1.784 |
| Scenario 1c | Beta(2,5) o F | 20 | 0.565 | 1.208 |
| Scenario 1c | Beta(2,5) o F | 40 | 0.643 | 1.53 |
| Scenario 1c | Beta(2,5) o F | 60 | 0.738 | 2.104 |

**Table 2 - Scenario 2**

The percentage of censored subjects and the estimated survival at the actual median survival time (S(ams)) are shown for various study lengths, where unit of time is ams (actual median survival).

| Study length | % censored | S(ams) |
|---|---|---|
| 1 * ams | 72.24 | 0.5 |
| 1.5 * ams | 62.18 | 0.5 |
| 2 * ams | 54.08 | 0.5 |
| 3 * ams | 42.15 | 0.5 |
| 4 * ams | 33.81 | 0.5 |
| 5 * ams | 27.99 | 0.5 |
| 6 * ams | 23.72 | 0.5 |

**Table 3 - Scenario 3**

The percentage of censored subjects and the estimated survival at the actual median survival time (S(ams)) are shown for various study lengths and recruitment periods, where unit of time is ams (actual median survival).

| Study length | Recruitment period | % censored | S(ams) |
|---|---|---|---|
| 3 * ams | 0.5 * ams | 14.95 | 0.5 |
| 3 * ams | 1 * ams | 18.04 | 0.5 |
| 3 * ams | 1.5 * ams | 21.99 | 0.5 |
| 3 * ams | 2 * ams | 27.06 | 0.5 |
| 4 * ams | 0.5 * ams | 7.48 | 0.5 |
| 4 * ams | 1 * ams | 9.03 | 0.5 |
| 4 * ams | 1.5 * ams | 11 | 0.5 |
| 4 * ams | 2 * ams | 13.52 | 0.5 |
| 5 * ams | 0.5 * ams | 3.74 | 0.5 |
| 5 * ams | 1 * ams | 4.52 | 0.5 |
| 5 * ams | 1.5 * ams | 5.5 | 0.5 |
| 5 * ams | 2 * ams | 6.76 | 0.5 |
| 6 * ams | 0.5 * ams | 1.87 | 0.5 |
| 6 * ams | 1 * ams | 2.26 | 0.5 |
| 6 * ams | 1.5 * ams | 2.75 | 0.5 |
| 6 * ams | 2 * ams | 3.39 | 0.5 |

**Table 4 - Scenario 4**

The percentage of censored subjects (administrative and dropout) and failures, as well as the estimated survival at the actual median survival time (S(ams)) and the ratio between the estimated median survival time and the actual median survival time (ms / ams) are shown for each simulation. Study length and the recruitment period are fixed and equal to 3 ams and 0.5 ams respectively, while we simulated three dropout distributions with P taking values 10%, 20% and 30%.

| Dropout distribution | % censored | S(ams) | ms / ams | % administrative | % dropout | % failures |
|---|---|---|---|---|---|---|
| Beta(5,2) o F | 23.47 | 0.516 | 1.047 | 13.54 | 9.93 | 76.53 |
| Beta(5,2) o F | 31.97 | 0.535 | 1.103 | 12.11 | 19.85 | 68.03 |
| Beta(5,2) o F | 40.48 | 0.557 | 1.168 | 10.7 | 29.78 | 59.52 |
| Uniform o F | 23.47 | 0.523 | 1.07 | 13.89 | 9.58 | 76.53 |
| Uniform o F | 31.97 | 0.55 | 1.153 | 12.82 | 19.15 | 68.03 |
| Uniform o F | 40.47 | 0.58 | 1.255 | 11.75 | 28.72 | 59.53 |
| Beta(2,5) o F | 23.47 | 0.531 | 1.094 | 14.2 | 9.26 | 76.53 |
| Beta(2,5) o F | 31.97 | 0.565 | 1.208 | 13.44 | 18.53 | 68.03 |
| Beta(2,5) o F | 40.48 | 0.602 | 1.35 | 12.69 | 27.79 | 59.52 |